\documentclass[11pt,a4paper,arial,onecolumn]{article}
\usepackage{latexsym}
\usepackage{amsmath}
\usepackage{amsfonts}
\usepackage{amsbsy}
\usepackage{amssymb}
\usepackage{epsfig}
\usepackage{mathrsfs}
\usepackage{epsfig}
\usepackage{psfrag}
\usepackage{bbm}

\usepackage[T1]{fontenc}
\usepackage[french]{babel}

\newcommand{\p}{\partial}

\usepackage{graphicx}

\DeclareGraphicsExtensions{.eps,.ps,.eps.gz,.ps.gz,.eps.Z}

\begin{document}


\begin{titlepage}
\begin{center}
\textsc{REPUBLIQUE TUNISIENNE}\\
\Large
\textsc\textbf{MINISTERE DU TRANSPORT ET DE L'EQUIPEMENT}\\
\textsc{Office de la Topographie et du Cadastre }\\

\vspace{3cm}
\LARGE
\textsc{\textbf{NOTE SUR LES REPRESENTATIONS QUASI-CONFORMES}}\\
[0.5\baselineskip]
Par\\[0.5\baselineskip]
\textbf{\textsc{Abdelmajid BEN HADJ SALEM}}\\
\vspace{0.5cm}
\normalsize \textsc{Ingénieur Général à l'Office de la Topographie et du Cadastre}\\
\normalsize \textsc{benhadjsalem@yahoo.co.uk}\\

\vspace{1cm}
\textsc{Mars 2011}
\\ 

\vspace{1cm}
\textsc{Version 1.}\\

\vspace{3cm}
\textsc{Office de la Topographie et du Cadastre \\ www.otc.nat.tn}

\end{center}
\end{titlepage}



\tableofcontents 



\clearpage
\Large
\begin{center}
\textsc{\textbf{    NOTE SUR LES REPRESENTATIONS QUASI-CONFORMES}} \\

\end{center}
\Large
\begin{center}
\textsc{\textbf{Abdelmajid BEN HADJ SALEM}}

\large
\textbf{Office de la Topographie et du Cadastre \\
BP 1056, 1080 Tunis Cedex}
\end{center}
\vspace{.5cm}
\normalsize
\section{Introduction}
En cartographie mathématique, on a étudié les représentations de la sphère avec les variables $(L_M,\lambda)$ ou celles de l'ellipsoide de révolution avec les coordonnées $(L,\lambda)$ vers le plan $(X,Y)$ avec:
\begin{align}
\left\{
\begin{array}{c}
        X=X(L_M,\lambda) \\
        Y=Y(L_M,\lambda)
\end{array}
\right.
\end{align}
ou:
 \begin{equation}
\left\{
\begin{array}{c}
        X=X(L,\lambda) \\
        Y=Y(L,\lambda)
\end{array}
\right.
\end{equation}
avec:
\begin{equation}
    L_M=\log \tan \left( \frac{\pi}{4}+\frac{\varphi}{2} \right) \quad \mbox{latitude de Mercator} \\
\end{equation}
et:
\begin{equation}
L=\log \tan \left( \frac{\pi}{4}+\frac{\varphi}{2} \right)-\frac{e}{2}\log \frac{1+e\sin\varphi}{1-e\sin\varphi} \quad \mbox{latitude isométrique}
\end{equation}
En posant:
\begin{equation}
    z=L_M+i\lambda
    \end{equation}
ou:
\begin{equation}
    z=L+i\lambda
    \end{equation}
\begin{equation}
    Z=X+iY
\end{equation}
on a considéré les représentations conformes (c'est-à-dire qui conservent les angles) ou encore définie par :
\begin{equation}
    Z=Z(z)
\end{equation}
avec $Z(z)$ une fonction dite holomorphe de $z$ soit:
\begin{equation}
\frac{\partial Z}{\partial \bar{z}}=0
\end{equation}
où $\bar{z}$ est le conjugé de $z$ soit $\bar{z}=L-i\lambda$.
\section{Les Représentations ou Transformations Quasi-Conformes}
\textbf{Définition:}\textit{ Une fonction $f(z)=Z=Z(z)$ définie et dérivable sur un domaine $\mathcal{D}\subset C $(l'ensemble des nombres complexes) est dite quasi-conforme si elle vérifie:
    \begin{equation}
\frac{  \partial Z}{\partial \bar{z}}=\mu(z).\frac{ \partial Z}{\partial z}
\end{equation}
avec:
\begin{equation}
     |\mu(z)|<1
\end{equation}}
$|z|$ désigne le module du nombre complexe $z$. Le coefficient $\mu$ s'appelle \textit{coefficient de Beltrami}.
\\
\subsection{Développement d'une fonction en un point $z_0$}
 Soit $f$ une fonction quasi-conforme et un point $z_0\in \mathcal{D}$. Ecrivons un développement de $f$ au point $z_0$. On a alors:
\begin{equation}
    f(z)=f(z_0)+(z-z_0)\frac{\p f}{\p z}(z_0)+(\bar{z}-\bar{z}_0)\frac{\partial f}{\partial \bar{z}}(\bar{z}_0)+...
\end{equation}
Par un changement de variable, on peut prendre $z_0=0$, d'où:
\begin{equation}
    f(z)=f(z_0)+z\frac{\partial f}{\partial z}(z_0)+\bar{z}\frac{\partial f}{\partial \bar{z}}(\bar{z}_0)+...
\end{equation}
Utilisant (10), l'équation précédente s'écrit en négligeant les termes du deuxième degré:
\begin{equation}
        f(z)=f(z_0)+z\frac{\partial f}{\partial z}(z_0)+\bar{z}\mu(z_0).\frac{\partial f}{\partial z}(z_0)
\end{equation}
Donc $f(z)$ s'écrit localement:
\begin{equation}
    f(z)=\alpha+\beta z+\gamma \bar{z}
    \end{equation}
où $\alpha,\beta,\gamma$ des constantes complexes avec:
\begin{equation}
     \left|\frac{\gamma}{\beta}\right|<1
\end{equation}
\section{Etude de la Transformée d'un cercle}
On sait que pour une transformation conforme, l'image d'un cercle autour d'un point est un cercle (ou encore l'indicatrice de Tissot est un cercle). Soit un point $z_0$ qu'on peut prendre égal à 0. Par un changement de l'origine des axes, la fonction $f$ s'écrit:
\begin{equation}
    f(z)=\beta z +\mu \beta \bar{z}
\end{equation}
Par abus, on garde la même notation. On considère autour de l'origine $z_0=0$ un point $M(x=a.\cos\theta,y=a.\sin\theta)$ qui décrit un cercle infiniment petit de rayon $a$. Etudions son image par $f$.
\\

De l'équation précédente, on a:
\begin{equation}
    z=a\cos\theta +ia\sin\theta =ae^{i\theta}
     \end{equation}

\begin{equation}
    \mu=|\mu|e^{ik}
\end{equation}
\begin{equation}
    \beta=|\beta|e^{il}
    \end{equation}

\begin{equation}
    f(z)=a|\beta|e^{il}(e^{i\theta}+|\mu|e^{i(k-\theta)})
\end{equation}
Si $\theta_1=\frac{k}{2}= \frac{\arg\mu}{2}$ \footnote{$z=x+iy=|z|e^{i\theta},\theta=\arg z$} on a: $z_1=ae^{ik/2}$ et :
\begin{equation}
    f(z_1)= a|\beta|e^{il}e^{ik/2}(1+|\mu|)
\end{equation}
\begin{equation}
    |f(z_1)|=a|\beta|(1+|\mu|)
\end{equation}
Maintenant prenons $\theta_2=\theta_1+\frac{\pi}{2}=\frac{k}{2}+\frac{\pi}{2}$, alors $z_2=ae^{i\theta_2}=ae^{ik/2}e^{i\pi/2}=iae^{ik/2}$ et on obtient:
\begin{equation}
        f(z_2)= ia|\beta|e^{il}e^{ik/2}(1-|\mu|)
           \end{equation}
         \begin{equation}
    |f(z_2)|=a|\beta|(1-|\mu|)
\end{equation}
en tenant compte que $|\mu|<1$.
\\

Des équations (21,23) et (25), on déduit que l'image de $M$ décrit une ellipse de grand-demi axe et demi-petit axe respectivement:
\begin{equation}
 a'=a|\beta|(1+|\mu|)
  \end{equation}
  \begin{equation}
 b'=a|\beta|(1-|\mu|)
\end{equation}
On appelle:
\begin{equation}
    K=\frac{1+|\mu|}{1-|\mu|}
\end{equation}
\textit{coefficient de distortion ou de dilatation}.

\section{Calcul d'un élément de longueur sur le Plan}
Un élément de longueur sur le plan est donné par:
\begin{equation}
    dS^2=dX^2+dY^2=|df|^2=df.\bar{df}
\end{equation}
Comme  $df=\beta dz+\gamma d\bar{z}$ et $d\bar{f}=\bar{\beta} d\bar{z}+\bar{\gamma} dz$, on alors:
\begin{equation}
    dS^2=dX^2+dY^2=|df|^2=df.\bar{df}=(\beta dz+\gamma d\bar{z})( \bar{\beta} d\bar{z}+\bar{\gamma} dz)  
    = \\ \beta\bar{\beta}dzd\bar{z}+\gamma\bar{\gamma}dzd \bar{z}+ dzd\bar{z} \left(\beta\bar{\gamma}\frac{dz}{d\bar{z}}+\gamma \bar{\beta}\frac{d\bar{z}}{dz} \right)  \qquad
    \end{equation}

 Posons:
\begin{equation}
    ds^2=dz.d\bar{z}
\end{equation}
Le carré du module linéaire de la transformation quasi-conforme s'écrit:
\begin{equation}
    m^2=\frac{dS^2}{ds^2}=|\beta|^2+|\gamma|^2+\left(\beta\bar{\gamma}\frac{dz}{d\bar{z}}+\gamma \bar{\beta}\frac{d\bar{z}}{dz} \right)
\end{equation}
Dans l'équation (32), considérons $z=ae^{i\theta}$ varie le long d'un cercle de rayon $a$ infiniment petit et faisons tendre $\theta \longrightarrow 2\pi$. Alors, on obtient :
\begin{equation}
    \frac{dz}{d\bar{z}}=\frac{aie^{i\theta} d\theta}{-aie^{-i\theta}d\theta}=-e^{2i\theta}=-1
      \end{equation}
         \begin{equation}
        \frac{d\bar{z}}{dz}=-e^{-2i\theta} =-1
\end{equation}
L'équation (32) devient:
\begin{equation}
    m^2=\frac{dS^2}{ds^2}=|\beta|^2+|\gamma|^2-\left(\beta\bar{\gamma}+\gamma \bar{\beta} \right)
\end{equation}
Comme:
$$ \gamma=\mu \beta $$
 on obtient:
\begin{equation}
        m^2=\frac{dS^2}{ds^2}=|\beta|^2+|\beta|^2|\mu|^2-\left(\beta\bar{\beta}\bar{\mu}+\mu \beta \bar{\beta} \right)
\end{equation}
or $\mu+\bar{\mu}=2|\mu| \cos\arg\mu $, par suite l'équation (36) s'écrit:
\begin{equation}
    m^2=\frac{dS^2}{ds^2}=|\beta|^2(1+|\mu|^2-2|\mu|\cos\arg\mu)
\end{equation}
Remplaçant $\beta$ par $\frac{\partial f}{\partial z}(z_0)$, (37) devient:
\begin{equation}
    m^2=\frac{dS^2}{ds^2}=\left|\frac{\partial f}{\partial z}(z_0)\right|^2 \left(1+|\mu|^2-2|\mu|\cos\arg\mu\right)       \end{equation}
\section{Exemple de Transformation Quasi-conforme}
Lors de passage de coordonnées planes $(X,Y)_i$ d'un système géodésique $S_1$ à des coordonnées planes $(X',Y')_j$ dans un autre système géodésique $S_2$, on utilise souvent une transformation du type:
\begin{equation}
    X'=X_0+aX+bY
     \end{equation}
  \begin{equation}
    Y'=Y_0+cX+dY
\end{equation}
ou encore sous forme matricielle :
\begin{equation}
    \left(
\begin{array}{c}
    X' \\
    Y'
\end{array}\right)=\left(
\begin{array}{c}
    X_0 \\
    Y_0
\end{array}\right)+\left(
\begin{array}{cc}
    a & b \\
    c & d
\end{array}\right).\left(
\begin{array}{c}
    X \\
    Y
\end{array}\right)
\end{equation}
En posant $Z=X'+iY'$ et $z=X+iY$, on obtient:
\begin{equation}
    Z=(X_0+iY_0)+X(a+ic)+Y(b+id)
\end{equation}
Posons:
\begin{equation}
    Z_0=X_0+iY_0
    \end{equation}
Comme $ X=(z+\bar{z})/2$ et $Y=(z-\bar{z})/2i$, alors l'équation (42) s'écrit:
\begin{equation}
Z=Z_0+z\left(\frac{a+d}{2}+i\frac{c-b}{2}\right)+\bar{z}\left(\frac{a-d}{2}+i\frac{b+c}{2}\right)
\end{equation}
Posons:
\begin{equation}
    \beta=\frac{a+d}{2}+i\frac{c-b}{2}
    \end{equation}
         \begin{equation}
    \gamma=\frac{a-d}{2}+i\frac{b+c}{2}
\end{equation}
 Alors (44) s'écrit:
\begin{equation}
    Z=Z_0+\beta z+\gamma \bar{z}
\end{equation}
Pour quelles valeurs de $a,b,c,d$ la transformation (41) est quasi-conforme? En comparant (44) avec (15), il faut que $|\gamma|<|\beta|$ soit:
\begin{equation}
    |\gamma|<|\beta|\Rightarrow |\gamma|^2<|\beta|^2 \Rightarrow \frac{(a-d)^2+(b+c)^2}{4}< \frac{(a+d)^2+(c-b)^2}{4} \nonumber \\
    \Rightarrow ad-bc>0
\end{equation}
C'est-à-dire que le déterminant de la matrice (41) soit positif.

 \section{Références}

1. \textbf{W. Zeng, L.M. Lui, F. Luo, T.F. Chan, S.T. Yau, X.F. Gu}: Computing Quasiconformal Maps on Riemann surfaces using Discrete Curvature Flow. 23p. Mai 2010. ArXiv:1005.4648v1.
\\

2. \textbf{L. Bers}: Quasiconformal mappings, with applications to differential equations, function theory and topology. American Mathematical Society Bulletin, vol 83 , no 6, pp 1083-1100,1977.
\\


\bibliographystyle{alpha} 


\clearpage



\end{document}